%% file: Murga1.tex
\documentclass[
aps,%
12pt,%
final,%
notitlepage,%
oneside,%
onecolumn,%
nobibnotes,%
nofootinbib,%
superscriptaddress,%
noshowpacs,%
centertags,%
natbib
]{revtex4} 

\usepackage{amsmath}
\usepackage{epstopdf}
\usepackage[usenames]{color}


\begin{document}
\selectlanguage{english}

\title{Restructuring and destruction of hydrocarbon dust in the interstellar medium}

\author{\firstname{M.~S.}~\surname{Murga}}
\email[E-mail: ]{murga@inasan.ru} 
\affiliation{Institute of Astronomy, Russian Academy of Sciences,
ul. Pyatnitskaya 48, Moscow, 119017 Russia}

\author{\firstname{S.~A.}~\surname{Khoperskov}}
\affiliation{Institute of Astronomy, Russian Academy of Sciences,
ul. Pyatnitskaya 48, Moscow, 119017 Russia}
\affiliation{Universit\'{a} degli Studi di Milano,
Dipartimento di Fisica, via Celoria 16, I-20133 Milano, Italy}
\affiliation{Sternberg Astronomical Institute, Lomonosov Moscow State
University, Universitetskii pr. 13, 119992 Moscow, Russia}

\author{\firstname{D.~S.}~\surname{Wiebe}}
\affiliation{Institute of Astronomy, Russian Academy of Sciences,
ul. Pyatnitskaya 48, Moscow, 119017 Russia}

\begin{abstract}
A model of key processes influencing the evolution of a
hydrocarbon grain of an arbitrary size under astrophysical conditions
corresponding to ionized hydrogen regions (HII regions) and supernova
remnants is presented. The considered processes include aromatization and 
photodestruction, sputtering by electrons and ions, and shattering due to
collisions between grains. The model can be used to simulate the
grain size distribution and the aromatization degree during
the evolution of HII regions and supernova remnants for a specified
radiation field, relative velocity of gas and dust, etc. The
contribution of various processes to the evolution of hydrocarbon dust
grains for parameters typical for the interstellar medium of our Galaxy
is presented. Small grains (less than 50 carbon atoms) should be
fully aromatized in the general interstellar medium. If larger grains initially
have an aliphatic structure, it is preserved to a substantial extent.
Variations in the size distribution of the grains due to their mutual collisions
depend appreciably on the adopted initial size distribution. For
the MRN initial distribution a significant redistribution of grain sizes
is obtained, which increases the mass fraction of smaller grains. 
Characteristic for an initial distribution from the work of Jones et al. 
(2013), with high initial fraction of small grains, is a general decrease in the
number of grains of all sizes.
\end{abstract}

\maketitle

\section{INTRODUCTION}

Interstellar dust manifests itself as a source of infrared (IR) and submillimeter
radiation, as well as a source of absorption in the optical and ultraviolet
(UV). One of the most prominent features of the dust emission spectrum is emission
bands at wavelengths from 2--20~$\mu$m. The origin of these bands is
believed to be related to aromatic hydrocarbon compounds. However, it is still unclear
in which form they are present in the interstellar medium (ISM). Initially, they were
associated with polycyclic aromatic hydrocarbons (PAHs); i.e., planar
macromolecules consisting of benzene rings~[1, 2]. Recently, it has been also
suggested that PAHs or individual benzene rings could be included in a more 
complex and less ordered structure known as hydrogenated amorphous carbon
(\hbox{a-C:H})~[3, 4] or mixed aromatic-aliphatic organic nanoparticles~[5].
Such compounds may exist as PAHs in some cases, however, in 
general, they are not identical to PAHs in their properties (structure, shape,
density, heat capacity, absorption cross section, etc.). Each model has
its advantages and drawbacks, and the ultimate choice of some specific representation
of dust properties is still to be made~[6--8].

Whatever compounds are responsible for dust emission in IR bands, observations
indicate that their absolute abundance and their abundance relative to larger
dust grains vary from galaxy to galaxy~[9--11], within a single galaxy~[12, 
13], within individual HII regions~[14], and also change with time~[15, 16]. In 
other words, the observations hint that the evolution of 
small dust grains (1) differs from the evolution of large dust grains, and 
(2) is sensitive to parameters of the local ISM (metallicity, velocity 
field, temperature, etc.). In objects with extreme physical conditions 
(HII regions, supernova-driven shells), the 
destruction of dust is efficient~[17], while, for example, in dense, cool molecular
clouds the opposite process may occur, which is the growth of dust grains by 
accretion or coagulation~[18, 19].

Observations with the Spitzer and Herschel space telescopes have become an important 
source of information about the properties of the dust component in galaxies 
and individual star-forming regions (SFRs), in particular, in photometric
bands centered on wavelengths of 8~$\mu$m and 24~$\mu$m. It is generally assumed
that the flux in the 8-$\mu$m photometric band ($F_8$) characterizes emission 
of PAHs, since several bright aromatic emission bands fall into this range, while
the 24-$\mu$m flux ($F_{24}$) is associated with emission of larger dust
grains. According to a number of studies~[13, 20], the flux
ratio $F_8/F_{24}$ can be used as a measure of such an important parameter of
the dust component as the fraction of aromatic dust grains in the overall
mass of dust ($q_{\textrm{PAH}}$).

One of well known properties of the dust component emission in SFRs in the near- and mid-IR
is the correlation of $F_8/F_{24}$ with the metallicity of the gas, which
is usually parameterized through the oxygen abundance, $12+\textrm{log(O/H)}$, in studies of SFRs.
Specifically, $F_8/F_{24}$ is lower in objects with
lower metallicities (see, e.g.,~[21]). Various mechanisms have been proposed
to explain this correlation, associating the content of metals either with the
formation efficiency or the destruction efficiency of the dust grains 
responsible for the IR emission at 8 and 24~$\mu$m.

The hypothesis that the $F_8/F_{24}-12+\textrm{log(O/H)}$ correlation is
associated with different efficiencies of the \emph{formation\/} of 
aromatic compounds in stars with different metallicities is not consistent
with the local character of this correlation. Essentially, the ratio of
IR fluxes correlates with the metallicity of the gas in which aromatic
compounds are currently located, not with the metallicity of
stars in which these compounds presumably have been synthesized. However, the
hypothesis relating a higher rate of \emph{destruction\/} of aromatic compounds
in the low metallicity medium also contradicts to observations. It was shown in~[15, 
16] that in low metallicity SFRs the $F_8/F_{24}$ ratio grows with age,
rather than decreasing with age, as would be expected if the evolution of
aromatic compounds was dominated by their destruction. A decrease of this
ratio with age is observed only in SFRs having solar or higher metallicity.
This suggests that the evolution of small dust grains and macromolecules
in SFRs has a complex nature, which is determined by balances between
several processes.

Interest to the detailed evolution of dust grains is caused by at least
two factors. First, their IR emission is often used as a proxy for the star formation
rate. Complicated evolution of dust grains in SFRs could lead to incorrect 
interpretation of IR observations. Second, the evolution of small 
hydrocarbon dust grains attracts attention from astrochemical and astrobiological 
points of view, since it is connected to the survival of organic compounds 
in the ISM. Moreover, the synthesis of molecular hydrogen may proceed at
higher temperatures on surfaces of \hbox{a-C:H}\ dust grains than on
surfaces of ``ordinary'' dust grains~[22].

Studying various aspects of dust evolution requires a
model which would take into account both processes of dust formation, growth, 
and destruction and possible grain structural variations, which are reflected
in their IR emission. So far, these
processes have been considered separately (see, e.g.,~[23] and [24]). In
the current study, we present a generalized model for the evolution of
dust grains, responsible for emission in IR bands, under the action of
processes typical for the ISM, HII regions, and supernova
remnants. The model is based on the study~[4], which examines the idea that
interstellar carbon dust grains consist of amorphous hydrocarbon material,  
\hbox{a-C:H}, whose structure changes when it is subject to UV irradiation.
The ability of these grains to radiate in IR bands is determined by the
degree of their aromatization. The optical properties of \hbox{a-C:H}
material for various aromatization degrees and the method to calculate
the aliphatic to aromatic transition rate are taken from~[4]. Unlike the previous models~[4, 
23, 25], our model includes several evolutionary processes, namely, photodestruction 
with restructuring, destruction due to collisions with high energy ions
and electrons, and destruction due to collisions between the grains themselves. 
We do not consider dust formation and growth processes in
this version of the model, but plan to include them in the future.

One of the key processes included in the model is the restructuring of dust grains 
illuminated by UV radiation, which forces transition of dust grains
from an aliphatic state saturated in hydrogen to a hydrogen-poor aromatic
state~[4, 26]. This process (called ``aromatization'' for short)
has been often observed in laboratory experiments on Earth~[27, 28] and
in near-Earth space~[29]. Aromatization likely also occurs in the ISM, especially in the presence
of enhanced (relative to the average circumsolar background) UV 
radiation~[30--33]. Note that accretion of hydrogen atoms from the ISM
could also lead to the opposite transformation from aromatic to 
aliphatic skeleton~[34]. However, the current version of the
model is motivated by our whish to study dust in complexes of ionized
hydrogen, where high-energy collisions with atomic hydrogen lead
to the destruction of dust grains, rather than to the incorporation of atoms into
a grain structure.

The destruction of dust grains by high-energy particles (electrons, ions) 
plays an important role in medium which is dominated by high-velocity (${\gtrsim} 10$~km/s) 
large-scale gas motions or high temperatures (${\gtrsim} 10^{4}$~K).
A ``molecular'' approach to treating the interactions of a carbon lattice
with incident particles was developed in [25, 35, 36] for computations of
the PAH destruction rate. This approach was adapted for more complex
\hbox{a-C:H}\ dust grains in~[37], where the necessity of its use in
studies of dust grains with fewer than 1000 atoms is emphasized. For larger dust grains
a less detailed ``classical'' approach can be used, where the efficiency of atom knocking from the
grain surface is calculated~[38, 39].

High speed collisions between dust grains (${>}1$~km/s) lead to their
shattering, which changes the grain size distribution~[38, 40].
The efficiency of this process depends on the properties of turbulent and
systematic motions in the medium. Under certain conditions, the fraction of
small dust grains can be substantially increased due to the destruction of
larger dust grains~[41, 42].

In the ISM these various processes compete with each other, and the role of
each process depends on specific conditions. Here we present a model of the evolution 
of the abundance of small hydrocarbon dust in the ISM. Section~2 presents our
general approach to describing the evolution of dust. The processes
considered are described in detail in Section~3. We analyze the contribution
of various processes to the evolution of dust in Section~4. Finally, we
formulate our main conclusions in 
Section~5.

\section{GENERAL DESCRIPTION OF THE DUST EVOLUTION}

We consider the evolution of \hbox{a-C:H} hydrocarbon dust grains,
which initially have an amorphous structure but can be partially
aromatized under the action of UV radiation and collisions with gas
particles (electrons, ions). There is no separate population of PAHs in
the model; they are represented by dust grains with a high aromatization
degree. Note that the model
can be used to study other dust populations, for example,
fully aromatized dust, with appropriate initial conditions. 

We used a discrete description of the dust grain population, assuming
that their initial size distribution can be described by some arbitrary
function $n(a)$, which corresponds to the mass distribution $\tilde{n}(m)$. 
Shattering due to grain-grain collisions produces
dust grains of various masses, causing the function $\tilde{n}(m)$ to change
with time. When one uses the distribution of dust grain number
per unit mass interval and a limited number of size (mass) bins, the
mass conservation can be violated in real computations, for example, when
a grain loses only a small amount of mass and remains in the same
mass bin, while the associated dust-grain fragment falls into a bin
corresponding to a smaller mass. To avoid violation of the mass conservation,
we introduced a mass-density function $\tilde{\rho}(m)$, which
specifies the total mass of dust grains per unit volume for a mass bin
$i$~[41]:
\begin{gather}
\tilde{\rho}_{i} = \bar{m}_i \tilde{n}_0(\bar{m}_i)(m_i-m_{i-1}),
\end{gather}
where $m_i$ and $m_{i-1}$ are the bin boundary values and $\bar{m}_i$ 
is the mean mass of the dust grains in the bin $i$:
\begin{gather*}
\bar{m}_i={\int\limits_{m_{i-1}}^{m_i}\tilde{n}_0(m)mdm\over\int\limits_{m_{i-1}}^{m_i}mdm}.
\end{gather*}
Here the subscript $0$ denotes the distribution function at the
initial time moment. Note that this is not the only way to do this. Violation of
mass conservations can also be avoided by normalizing variations in
a mass bin to the mean mass in the bin (see [43] for more detail).

A key difference of our model from other similar studies is that we 
considered dust grains not only of different masses, but also of different
aromatization degrees. It is assumed that the degree of aromatization 
of the dust grains can be described by a wide energy bandgap for the dust material
$E_{\textrm{gap}}$, which reaches a maximum when the dust is fully
hydrogenated ($E_{\textrm{gap}}\sim3$~eV) and approaches zero in the
aromatic state. Therefore, in our model, the mass-density function
$\tilde{\rho}$ is a function of two parameters: the mass of dust grains 
$m$ and the bandgap energy $E_{\textrm{gap}}$. The number of size (mass) bins is
$N^{\textrm{a}}$, the number of bins over the bandgap is 
$N^{\textrm{eg}}$. The mass distribution of the dust grains allows for various
descriptions and normalization. In this study, we characterized the mass
using the total number of carbon atoms in the grain, $N_{\textrm{C}}$.

The time variation of the quantity $\tilde{\rho}_{ij}$ in the mass bin $i$ and 
the energy bandgap bin $j$ is the sum of the variations due to each of the 
processes included in the model:
\begin{gather} \label{main_difeq}
\frac{d\tilde{\rho}_{ij}}{dt} =
\left[\frac{d\tilde{\rho}_{ij}}{dt} \right]_{\textrm{arom}}+
\left[\frac{d\tilde{\rho}_{ij}}{dt} \right]_{\textrm{photo}}+\\
\nonumber{}+ \left[\frac{d\tilde{\rho}_{ij}}{dt}
\right]_{\textrm{sput}}+ \left[\frac{d\tilde{\rho}_{ij}}{dt}
\right]_{\textrm{shat}},
\end{gather}
where the subscripts ``arom'', ``photo'', ``sput,'' and ``shat'' refer to the
processes of aromatization, photodestruction, sputtering, and shattering,
respectively. The process of aromatization changes the distribution of the
dust grains only over $E_{\textrm{gap}}$, photodestruction changes their
mass distribution, and the remaining processes affect the distributions
over both of these parameters. Here, we do not follow the mass-density
function per se, but rather variations in the number of hydrogen atoms
in the dust grain and the degree of its aromatization. We assumed that all the
dust grains in a bin $ij$ evolve in the same way. If variations in
$N_{\textrm{C}}$ and $E_{\textrm{gap}}$ in a bin give rise to new values
of these parameters that correspond to a different bin, all the dust grains
are moved to that other bin. In other words, if the parameters of a bin
have changed only insignificantly and remain within the limits for that bin,
the mass density of the bin does not change. If the parameters of a bin
have changed significantly, the bin with the corresponding values of
$N_{\textrm{C}}$ and $E_{\textrm{gap}}$ is found, and that bin acquires 
the mass density of bin $ij$. When considering the process of shattering,
the mass-density function was computed directly.

The next section describes the computation of the individual terms in the
set of equations~(2). The results are presented in the form of the rates
$R$ corresponding to the numbers of acts of specific processes per unit time
per dust grain.

\section{MICROSCOPIC PROCESSES}

A dust grain in the ISM is gradually affected by various external influences,
which can lead to changes in both its size and its structure. It can lose
some of its atoms, fragment into smaller dust grains or into atoms and
simple chemical compounds. These variations are mainly due to absorption
of photons, as well as due to collisions of dust grains with energetic particles and
with each other. These processes have been considered in the literature
earlier, but some of them have been studied only in relation to PAHs. Here,
we present a corrected approach suitable for ``generalized'' \hbox{a-C:H} dust 
grains. Note that the smallest of the dust grains we have considered are
essentially macromolecules. In the literature, the term macromolecule is
sometimes used to refer to PAHs. In our model, there is no such thing as
a ``pure'' PAH, since PAHs are a subset of \hbox{a-C:H}\ dust with a
continuous size distribution. Therefore, to avoid confusion, we will refer
to all the dust particles we have considered as grains. At the same time, for some
processes associated with the smallest grains, we apply theories 
developed for macromolecules, together with the corresponding terminology.

\subsection{Interaction with Radiation}

Photons with energies $E>3$~eV, i.e., UV photons, exert a significant
influence on small dust grains and macromolecules. The absorption of
such a photon by a dust grain can have several outcomes --- dissociation,
radiation, and ionization. The probability that one of these outcomes
is realized depends on its activation energy, the photon energy, and the
size and composition of the dust grains. In its general form, the rate
$R$ (in units of events per second) for any photoprocess occurring in
a grain with radius $a$ under the action of a radiation field $F(E)$ is
\begin{gather}
R(a) = \int\limits_0^{\infty} Y(a,E) C_{\textrm{abs}}(a,E) F(E)
dE. \label{commonR}
\end{gather}
Here, $F(E)$ is the radiation flux written in terms of the number of photons 
with energy $E$, $Y$ is the probability of occurrence (yield) of a process 
upon absorption of a photon with energy~$E$, and $C_{\textrm{abs}}(a,E)$ is
the grain absorption cross section. In our computations we used the optical properties
of \hbox{a-C:H} calculated in~[44--46]. Further, 
Eq.~(3) was used to take into account the contribution of photoprocesses
in the set of equations (2).

The probabilities of breaking a C--H bond and of photoionization are known
from experiments and/or theoretical computations (see below). The probabilities
for the other processes can be found if the rate constants $k_i$ for all
possible processes are known; in our case, this includes radiation, 
ionization, and breaking of C--C bonds (a sum over these processes is in the
denominator, over the subscript $j$). The probability of the occurrence of
process $i$ can be found using the formula
\begin{gather}
Y_i = \frac{k_i}{\sum\limits_{j=1,3} k_j}. \label{eq:yield}
\end{gather}
Here, we take the rate constant $k_i$ of a process to be its rate in units
of events per second, subject to the condition that the probability of 
obtaining some result is equal to unity.

\textbf{3.1.1. Aromatization and photodissociation.} When modeling 
photodissociation, we adopted the general scheme presented in~[47] for
PAHs, extending it to \hbox{a-C:H} dust grains. In this scheme, the
partial destruction of \hbox{a-C:H} dust grains can be divided into two
stages. First, the main channel is the removal of a hydrogen atom,
(dehydrogenation). The bonds with neighboring carbon atoms that are freed 
after the removal of the hydrogen atom form new bonds, which gradually form
a closed ring: the structure of the dust grains makes a transition from
an aliphatic to an aromatic state~[27, 28]. Laboratory studies of this
process are presented in~[26--28, 48], and theoretical computations of
its rate are described in~[49, 50].

Initially, the value of $N_{\textrm{C}}$
corresponding to the mean mass for a given bin and the constant value
$E_{\textrm{gap}}=2.67$~eV is assigned to each mass interval. We used
the relation $E_{\textrm{gap}}=4.3X_{\textrm{H}}$~[49] to estimate the
number of hydrogen atoms in a grain, where
\begin{gather*}
X_{\textrm{H}}={N_{\textrm{H}}\over
N_{\textrm{C}}+N_{\textrm{H}}}.
\end{gather*}
During the evolution of the grain, the fraction of C--H bonds
($X_{\textrm{H}}$) varies, as well as the fraction of aliphatic
($X_{{\textrm{sp}}^3}$) and aromatic ($X_{{\textrm{sp}}^2}$) C--C bonds. 
Expressions for calculating $X_{{\textrm{sp}}^3}$ and $X_{{\textrm{sp}}^2}$ 
can be found in~[44].

At the \emph{first stage}, the main channel for dissociation is a loss of a
hydrogen atom. The cross section for this reaction,
$\sigma^{\textrm{CH}}_{\textrm{loss}}$, has been measured in a number of
experiments~[26, 51, 52] for radiation energies from ${\sim}6$ to 
${\sim}22$~eV, and varies from $10^{-21}$~cm$^{-2}$ to $1.5\times 
10^{-19}$~cm$^{-2}$. We used the mean value 
$\sigma^{\textrm{CH}}_{\textrm{loss}}=10^{-20}$~cm$^{-2}$~[45] for all 
photons with energies from 10 to 13.6~eV, and adopted a cross section of
zero outside of this range. The probability of dissociation, 
$Y_{\textrm{diss}}^{\textrm{CH}}$, is about 0.1~[53]. The rate at which
C--H bonds are broken (and, in our model, the rate of aromatization) per
dust grain is
\begin{gather}\label{eq:photorates}
R_{\textrm{arom}}(a,E) =\\
\nonumber{}= \int\limits_{10\textrm{~eV}}^{13.6\textrm{~eV}}
Y_{\textrm{diss}}^{\textrm{CH}}{\sigma^{\textrm{CH}}_{\textrm{loss}}\over
\pi a^2} C_{\textrm{abs}}(a,E) F(E) dE.
\end{gather}
This expression is based on formula~(32) from~[53]. The upper limit of the
integration, 13.6~eV, is justified by the fact that we consider the 
evolution of dust grains in an ``ordinary'' ISM, where shorter-wavelength
photons are absent. In the vicinity of hot stars, it is necessary to take into account more energetic radiation, and consequently the cross section
$\sigma^{\textrm{CH}}_{\textrm{loss}}$ for higher energies.

The model assumes that the number of broken bonds is limited by the depth
to which a photon penetrates into the dust grain. The maximum penetration
depth is assumed to be 200~\AA\ for all dust grains, independent of the
photon energy~[45]. This means that small dust grains will be aromatized
throughout their volume, while grains with radii exceeding 200~\AA\ will be
aromatized only in a surface layer.

When the fraction of hydrogen atoms $X_{\textrm{H}}$ becomes less than
$5\%$, the \emph{second stage} ensues, when another dissociation channel
is switched on~[54, 55] --- removal of \hbox{C$_2$}. It is
suggested in~[47] that PAH macromolecules eject acetylene molecules upon
photodissociation, but \hbox{C$_2$} will likely be removed from
fully dehydrogenated dust grains.

Essentially, destruction via the loss of \hbox{C$_2$}\ is efficient only
for small dust grains or macromolecules. In particular, it was shown 
in~[47, 56] that the second stage plays an important role only for
macromolecules with less than 50 atoms. The main dissociation channel
for larger dust grains is removal of hydrogen atoms~[57]. The value of
$X_{\textrm{H}}$ required for the onset of the second stage is not reached
by such grains.

As in [47], we used the Rice-Ramsperger-Kassel (RRK) approximation~[58] 
adapted for PAHs in [56] to calculate the rates of the removal of an atom
or a group of atoms. In this approximation it is assumed that the rate at which
an element is removed from a macromolecule with a given number of atoms
$N_{\textrm{atom}}$ is related to its internal energy $E_{\textrm{int}}$:
\begin{gather}
k_{\textrm{loss}} =
k_0(1-E_0/E_{\textrm{int}})^{3N_{\textrm{atom}}-7} \label{kloss}
\end{gather}
with $k_0 = (10^{16},10^{16},10^{15})$~s$^{-1}$ for H, H$_2$, and C$_2$, 
respectively. It is assumed that the internal energy of a dust grain 
$E_{\textrm{int}}$ is equal to the energy of the absorbed photon $E$. Unlike
[47, 56], where the region of neutral hydrogen with 
$E<13.6$~eV was considered, we extrapolate this expression to higher
energies.

The value of $E_0$ for several small PAHs was obtained experimentally~[56], 
however, it was shown in~[35] that this experimental value is not quite
appropriate for a mixture of interstellar PAHs, and several values of $E_0$
were used instead. In our model, this quantity is considered as a parameter.

\textbf{3.1.2. IR emission.} The rate at which IR photons are emitted
$k_{\textrm{IR}}$ can be estimated in terms of the absorption cross section
$C_{\textrm{abs}}$~[59]:
\begin{gather}
k_{\textrm{IR}}(E) = \int{\frac{C_{\textrm{abs}}(\lambda) 4\pi
B_{\lambda}(T\left[E\right]) d\lambda}{h c/\lambda}}, \label{kir}
\end{gather}
where $T\left[E\right]$ is the vibrational temperature corresponding to the
energy of the absorbed photon $E$, $B_{\lambda}(T\left[E\right])$ is the
Planck function corresponding to the temperature $T\left[E\right]$ at
wavelength $\lambda$, and $h$ and $c$ are the Planck constant and the speed of
light.  The temperature $T\left[E\right]$ can be found if the specific heat
capacity of the material $C_V$ is known, which for \hbox{a-C:H}\ is the
sum of the specific heat capacities of the C--H ($sp^3$) and C--C ($sp^2$ 
and $sp^3$) bonds [4], in proportions corresponding to the relative numbers
of these two types of bonds
\begin{gather}
C_V   = \frac{1}{1-X_{\textrm{H}}} \Big[ X_{\textrm{H}}C_V({\textrm{CH}})
+\\
\nonumber{}+X_{{\textrm{sp}}^2}C_V({\textrm{CC}}_{{\textrm{sp}}^2})+
X_{{\textrm{sp}}^3}C_V({\textrm{CC}}_{{\textrm{sp}}^3})\Big].
\end{gather}
We applied this formula to calculate the specific heat capacity, using the
technique of~[60] for the C--C and C--H aromatic bonds and the experimental
data for polyethylene~[61] for aliphatic bonds. The derived heat
capacities demonstrate a tendency to grow as the fraction of hydrogen
atoms or the bandgap increases~[62], so that the temperatures of dust 
grains of the same internal energy can differ appreciably for hydrogenated
and dehydrogenated amorphous hydrocarbon material.

\textbf{3.1.3. Ionization.} The ionization probability function  
$Y_{\textrm{ion}}^{\textrm{V}}$ from [63] was utilized for PAHs in~[47], which
we adopt as a basis for calculating the rates of various destruction processes.
According to this expression, all the energy goes into ionization when 
the photon energy is $E>17$~eV. Since photons with such energies were not 
considered in [63], this choice of an ionization probability function
was not of utmost importance for that study. Since we wish to apply our 
model to studies of the evolution of dust inside HII regions, we instead used
the ionization probability function $Y_{\textrm{ion}}^{\textrm{WD}}$ proposed
for PAHs in [64] to estimate the probability of ionization of the dust grain
upon absorption of a photon. This expression makes it possible to model the
response of the dust grain to absorption of a hard photon in more detail,
admitting both ionization and destruction of the dust grains.

Since photons with energies higher than 17~eV are present only in the 
vicinities of massive stars, our choice of the ionization probability function
is important only for estimating photodestruction rates near such objects.
However, in our model, the destruction probability function is also used
to calculate the rate of destruction of dust grains due to collisions with high-velocity
ions and electrons, which can transfer a substantial energy to a grain, 
making our choice of $Y_{\textrm{ion}}$ more critical.

\subsection{Interactions with Gas}

Ions and electrons can also possess kinetic energies sufficient to
destroy dust grains with which they collide. The behavior of this
interaction differs qualitatively for large and
small ($N_{\textrm{C}}<1000$) dust grains, which essentially correspond
to large molecules. Large dust grains are bombarded by energetic particles
only at their surfaces, and are affected only by surface sputtering, while the
bombardment of small dust grains can occur throughout their volume, leading
to their complete destruction. Two types of interactions between energetic
particles and hydrocarbon dust grains can be distinguished: 1)~removal of
a hydrogen atom or \hbox{C$_2$}\ molecule upon excitation of a dust grain, 
2)~direct removal of an atom from the lattice due to a collision. A
description of the interactions of small and large dust grains with gas
adopted in our model is presented below.

\textbf{3.2.1. Statistical sputtering of small dust grains.} Here, 
statistical sputtering refers to the dissociation (removal) of a structure
element of a dust grain as a result of de-excitation following the
absorption of the energy of an incident particle when it interacts with
the electron cloud of the dust grain. This type of collision is also
referred to as an inelastic collision. As a whole, the situation is
similar to the absorption of a photon, but in this case, the kinetic energy
of the incident particle can be transferred to the lattice only partially.
Only charged particles exert a substantial influence in statistical 
sputtering; i.e., only ions and electrons. Neutral atoms do not
interact with the electron cloud of a dust grain, and can act on the
lattice only when they impact one of its nodes directly.

The transferred energy $E_{\textrm{tr}}$ depends on the efficiency of
the friction experienced as the particle moves inside the dust grain
and the distance traversed. Computations of these quantities were carried 
out for PAHs in~[35]. Here, we used analogous computations with changes
related to the different geometry of the dust grains considered. We
assume that the molecular structure of an \hbox{a-C:H} dust grain can be
represented by a cube consisting of $N_p$ layers, each of which is
a quasi-PAH~[37] 
(of course, real \hbox{a-C:H} dust grains are unlikely to have such a simple
shape). In this geometry, the total number of layers in the
grain can be estimated as $\sqrt[3]{N_{\textrm{C}}-1}$. Each layer
contains  $N_{\textrm{C}}^{p} = (N_{\textrm{C}})^{2/3}$ atoms. Passing
through one layer, an ion transfers an energy $E_{\textrm{tr}}^{0}$ to the
grain, which can be calculated as described in Appendix~A. The total energy
transferred by the ion to the \hbox{a-C:H} grain, $E_{\textrm{tr}}$, will
be $N_p \times E_{\textrm{tr}}^{0}$ if the ion passes through the whole grain,
and $N_l \times E_{\textrm{tr}}^{0}$ if the number of layers
traversed by the ion $N_l=E_{\textrm{ion}}/E_{\textrm{tr}}^{0}$ is less
than $N_p$.

The processes occurring in the grain as a result of such absorption of
energy were treated analogously to the absorption of a photon, with 
$E_{\textrm{int}}$ in (6) replaced by $E_{\textrm{tr}}$. If the dust
grains are hydrogenated, this energy (taking into account the dissociation
probability $Y^{\textrm{CH}}_{\textrm{diss}}$) will go into the removal
of hydrogen atoms and aromatization. In the case of dehydrogenated grains,
dissociation occurs via the loss of C$_2$ molecules.

Taking into account all possible arrival directions of the ions (the 
angle $\theta$), the rate at which dust grains are destroyed as a result 
of statistical sputtering by ions $R_{\textrm{s}}^{\textrm{ion}}$ can
be calculated using the formula
\begin{gather}
R_{\textrm{s}}^{\textrm{ion}} = v_{\textrm{ion}} n_{\textrm{ion}}
\int\limits_{0}^{\pi/2} \sigma_g
Y\big[E_{\textrm{tr}}(\theta)\big] \sin\theta d\theta,
\label{rate_electron}
\end{gather}
where $n_{\textrm{ion}}$ is the number density of ions in the medium and 
$\sigma_g$ is the geometrical cross section of the \hbox{a-C:H} dust grains. 
This must be multiplied by two to take into account destruction due to the
removal of \hbox{C$_2$}\ groups. Note that, when considering
interactions of an ion with a dust grain, the charge of the grain must also
be taken into account in the Coulomb factor, but we have not concerned
ourselves with this issue, since we assume that all the dust grains are
neutral. In fact, the grain charge can substantially influence some 
microprocesses; for example, it can play an important role in the
dynamics of dust in HII regions and shocks~[65, 66], by determining the
relative velocity of the dust grains and gas. However, in our case, 
introducing a grain charge would mean considering a three-dimensional rather 
than two-dimensional distribution of the dust grains, which appreciably
complicated the computations. Since the collisional velocities in this
study are specified and not derived from a physical model, the only
error introduced by assuming the dust grains are neutral is associated
with the Coulomb factor.

Formula (9) can be used for a specific value of velocity of an incident ion
$v_{\textrm{ion}}$, for example, due to non-thermal gas motions. When
thermal motion of the ions is considered, we must integrate over all possible
velocities:
\begin{gather}
R_{s,T}^{\textrm{ion}} = 2
\int\limits_{v_{\textrm{ion}}^0}^{\infty} R_{s}^{\textrm{ion}}
 f(v_{\textrm{ion}}, T_{\textrm{gas}}) d v_{\textrm{ion}},
\label{eq: thermal rate}
\end{gather}
where $f(v_{\textrm{ion}}, T_{\textrm{gas}})$ is a Maxwellian distribution
for given ions at a temperature $T_{\textrm{gas}}$ and $v_{\textrm{ion}}^0$
is the speed corresponding to the minimum energy required for dissociation,
estimated using (9).

Incident electrons can also interact with the electron cloud of a dust grain.
The electrons in a hot gas can reach kinetic energies of several tens or
even hundreds of eV, which can lead to the destruction of a dust
grain during a collision. To estimate the energy $E_{\textrm{tr}}^e$ obtained
from an electron, one needs to calculate the Mott cross section~[67], which is
a laborious task. It was proposed in [35] to use the experimental approximation
from [68], obtained for graphite, for this cross section. We used the
algorithm of [35] to compute $E_{\textrm{tr}}^e$, which is taken to be
the difference between the initial energy of the electron $E_{0}^{e}$ and 
its energy after it has passed through a PAH, $E_{1}^{e}$, which is
found taking into account the traversed distance. This distance was estimated
in the same way as that for an ion (Eq.~A2). The energy is also multiplied
by a factor of $N_l$ to take into account the volume of the \hbox{a-C:H}
dust grain. We then calculated the probability of the removal of a
\hbox{C$_2$}\ group and found the rate of destruction of dust grains by
electrons using formula (10), replacing the subscript and superscript 
``ion'' with ``e''. Due to the low mass of the electron, with typical
velocities for large-scale motions in the ISM, the influence of non-thermal
collisions is negligible, and we have accordingly taken into account only
thermal collisions.

\textbf{3.2.2. Non-statistical sputtering of small dust grains.} 
Non-statistical sputtering refers to the removal of atoms from the
lattice of dust grains as a result of interactions between the nuclei of
atoms in the lattice and an incident ion with a kinetic energy exceeding 
the binding energy of the atoms to the lattice.

In the computations of the destruction efficiency in [25], elastic
collisions between particles (hydrogen, helium, carbon) moving with
a velocity $v_{\textrm{ion}}$ and stationary carbon atoms in the lattice
are considered. As a result of a collision, some fraction of the kinetic energy
of the incident particle is transferred to a carbon atom. If the acquired
energy exceeds the critical energy for removing the atom from the lattice
($E_{\textrm{crit}}$), the dust grain will lose an atom. The value of
$E_{\textrm{crit}}$ can be estimated only approximately, first, due to the
lack of reliable experimental data and second, because this quantity most
likely depends on the size, structure, and charge of the dust grain, as 
well as the position of the atom in the lattice: atoms at the edge of the
grain can be removed much more easily than atoms in the central part of the
grain. In our model, we adopted the value $E_{\textrm{crit}} = 7.5$~eV, 
as was done in~[25] for PAHs. Generally speaking, we consider not
PAHs, but rather compounds that could include both aliphatic and aromatic 
bonds. Since aliphatic bonds are not as strong as aromatic bonds, the
kinetic-energy threshold could be lower for the former. However, as a rule,
a molecular treatment of the bombardment process is only important for very
small dust grains, which rapidly aromatize, as was noted above. Accordingly,
their bonds are to a large extent aromatic.

Such collisions are described in [25] using the Ziegler-Biersack-Littmark
(ZBL) theory [69], which can be used to find the energy 
$\langle E_{\textrm{tr}}^{\textrm{elas}}\rangle$, transferred to a carbon
atom by an incident particle, as well as the interaction cross section
$\sigma$. The main formulas are presented in Appendix~B. Further, we can
calculate the collision rate for a single atom in the lattice. For a flux
of ions moving perpendicular to the lattice, this collision
rate can be calculated as
\begin{gather}
R_{n,0}^{\textrm{ion}} = n_{\textrm{ion}} v_{\textrm{ion}}\sigma.
\label{rate_nuclear}
\end{gather}

To calculate the collision rate for the dust grain as a whole, we must
multiply the rate $R_{n,0}^{\textrm{ion}}$ by the number of possible
collisions. Because multiple collisions between an incident particle and atoms
in the lattice are possible in bulky grains, the destruction rate can be
increased appreciably compared to the destruction rate for a flat structure
(PAH). If the energy of the incident ion is sufficient to knock out an atom
from the first layer, its energy decreases by 
$\langle E_{\textrm{tr}}^{\textrm{elas}}\rangle$. Our calculation of the
energy $E_{\textrm{tr}}^{\textrm{max}}$ transferred to the atom in a 
head-on collision is described in Appendix~B. The quantity
$\langle E_{\textrm{tr}}^{\textrm{elas}}\rangle$ is the energy 
$E_{\textrm{tr}}^{\textrm{max}}$ averaged over all possible arrival angles 
for the ion. In the second layer, the ion carries the energy 
$E_0^{\textrm{ion}}-\langle E_{tr}^{\textrm{elas}}\rangle$. If this residual
energy is again sufficient to knock out atom, the second layer will also
lose an atom from its lattice. This process will continue until the ion
leaves the dust grain or its remaining kinetic energy becomes too small to remove
another atom. The number of collisions leading to the removal of an atom
$N_{\textrm{col}}$ can be estimated as ${\rm min}(E_0^{\textrm{ion}}/\langle
E_{\textrm{tr}}^{\textrm{elas}}\rangle, N_p)$. The rate of destruction of
\hbox{a-C:H} dust grains is then given by 
\begin{gather}
R_n^{\textrm{ion}}(v_{\textrm{ion}}) = N_{\textrm{col}}
N_{\textrm{C}}^{p} R_{n,0}^{\textrm{ion}}.
\end{gather}
In the case of thermal motion of the gas, we must integrate this destruction
rate over the full range of velocities, as in the case of statistical
destruction [Eq.~(10)]. As was already noted, the most abundant ions
are ions of hydrogen and helium; therefore, we consider the destruction 
caused by these ions for both statistical and non-statistical sputtering,
together with carbon ions being representative of heavier ions.

\textbf{3.2.3. Sputtering of large dust grains.} As was shown in [37], 
the interactions between the lattice and incident particles does not
need to be considered in detail in the case of large dust grains 
($N_{\textrm{C}}>1000$), so that we can restrict our treatment to a classical
description of this process. If an atom at the surface of a dust grain 
acquires sufficient energy during a collision with an incident particle
to break its bonds, it leaves the grain. The efficiency of this
process has been calculated in multiple studies, for several astrophysically
important materials, including graphite and amorphous carbon [38, 40, 70]. 
The probability of ejecting an atom from the surface
$Y_{\textrm{sput}}$ for \hbox{a-C:H} dust grains was calculated in [39].
It was shown that the sputtering rate for \hbox{a-C:H}\ was higher
than for graphite due to its lower density and the deeper penetration of
the ions. It was also found that the degree of aromatization does not
influence the sputtering process. We used the $Y_{\textrm{sput}}$ value
obtained in~[39] for H$^{+}$ and He$^{+}$ using the Stopping and Range of 
Ions in Matter (SRIM) program~[69]). The rate of sputtering of dust grains 
with radius $a$ bombarded by ions with velocity $v_{\textrm{ion}}$ can be
calculated as
\begin{gather}
R_{\textrm{sput}}^{\textrm{nt}} = 2\pi a^2 \sum v_{\textrm{ion}}
n_{\textrm{ion}} Y_{\textrm{sput}}^{\textrm{ion}}.
\end{gather}
In the case of thermal sputtering, this rate is calculated for the entire range
of possible velocities [Eq.~(10)]. According to~[39], this rate should be
multiplied by four to take into account the influence of the arrival angle
of the ion.

\subsection{Shattering}

The last process we consider is collisions between dust grains that lead
to their fragmentation into smaller grains. Depending on the velocity of the
collision, a dust grain can be destroyed partially or completely. The size and
number of fragments must be calculated for each collision. A theoretical
description of this process and its application to a number of astrophysical
objects are presented in~[38, 40--42]. Below we summarize the main assumprion of
the theory used in our model.

It is assumed that a grain P with mass $m_{\textrm{P}}$ collides with another
grain T with mass $m_{\textrm{T}}$ ($m_{\textrm{P}}\le m_{\textrm{T}}$) with
the relative velocity of $v_{\textrm{col}}$. As a result of their collision, 
grain P is completely destroyed, regardless of its mass. The mass fraction
$M_{\textrm{sh}}/m_{\textrm{T}}$ of the dust grain T through which the shock 
generated by the collision passes can be estimated for a given collision
velocity and dust material using formulas from~[40, 41]. Like the
authors of those studies, we assumed that, if $M_{\textrm{sh}}$ comprises
more than half of the mass of the dust grains $m_{\textrm{T}}$, the grain
is completely destroyed. Otherwise, the fraction $M_{\textrm{shat}}=
0.4M_{\textrm{sh}}$ is fragmented into smaller dust grains. The fragments
of both dust grains have power-law size distributions:
\begin{gather}
n_{\textrm{shat}} = C a_{\textrm{shat}}^{-3.3} 
\end{gather}
where the factor $C$ can be found from the normalization.

To calculate the parameters of the forming fragments, we must take into
account all possible collisions between grains of various sizes and types.
In this study, we considered collisions between hydrocarbon dust grains of
various sizes and degrees of aromatization, excluding collisions with
silicate grains for the moment (their contribution will be considered in
future studies).

Our model comprises $N^{\textrm{a}}$ dust size bins and 
$N^{\textrm{eg}}$ bins of the bandgap $E_{\textrm{gap}}$; therefore,
$N^{\textrm{a}}\times N^{\textrm{eg}}$ types of collisions are possible.
The size (mass) distribution of the fragments is described above.
The distribution of the degree of aromatization of the fragments is estimated as
follows. If a dust grain has a radius smaller than 200~\AA, i.e., it is
uniform in terms of its degree of aromatization, its fragments 
remain in the same $E_{\textrm{gap}}$ bin as the initial dust grain.
If a dust grain has a radius of more than 200~\AA, is consists of two
layers --- an inner layer with the initial degree of aromatization and
an outer layer whose degree of aromatization corresponds to the current
$E_{\textrm{gap}}$ bin. If the mass fraction of the destroyed material is
lower than the mass fraction of the outer layer (with thickness 200~\AA),
the fragments are assigned the $E_{\textrm{gap}}$ value of the surface
layer. Otherwise, the fragments are distributed among the bins with the
$E_{\textrm{gap}}$ values for the inner and outer layers, assuming that
the outer layer was completely destroyed and the remaining fragments have
the $E_{\textrm{gap}}$ value of the inner layer.

The numerical rate of variation of the mass density of a bin as a result
of shattering of the dust grain can be expressed as
\begin{gather}\label{ro_frag}
\left({d\tilde{\rho}_{ij}\over dt}\right)_{\textrm{shat}}  =
-\bar{m}_i \tilde{\rho}_{ij} \sum\limits_{j_1=1}^{N^{\textrm{eg}}}
\sum\limits_{k_1=1}^{N^{\textrm{a}}}\alpha_{i k_1}
\tilde{\rho}_{k_1 j_1}
 +\\
 \nonumber{}+  \sum\limits_{j_1=1}^{N^{\textrm{eg}}}\sum\limits_{j_2=j_1}^{N^{\textrm{eg}}}\sum\limits_{k_1=1}^{N^{\textrm{a}}}\sum\limits_{k_2=k_1}^{N^{\textrm{a}}}\alpha_{k_1 k_2}\tilde{\rho}_{k_1 j_1} \tilde{\rho}_{k_2 j_2}m^{{\rm shat},i,j}_{k_1 j_1 k_2 j_2}
 \\ \nonumber
\alpha_{k_1 k_2}  =  \frac{\sigma_{k_1 k_2} v_{k_1
k_2}}{\bar{m}_{k_1} \bar{m}_{k_2}},
\end{gather}
where $m^{{\rm shat},i,j}_{k_1 j_1 k_2 j_2}$ is the mass of the fragments
falling in the bin ($i$,$j$) during a collision of dust grains from bins 
($k_1$,$j_1$) and ($k_2$,$j_2$) and $\sigma_{k_1 k_2}$ is the collision
cross section for dust grains with masses from bins $k_1$ and $k_2$
(determined as $\pi(a_{k_1}^2+a_{k_2}^2)$). The method used to calculate
$m^{\rm shat}$ was taken from~[41]. After taking into account all collisions,
the degree of aromatization in the bin is calculated as the mean-weighted
degree of aromatization of all fragments in the bin, where the mass density
acts as the weighting factor.

The process of aromatization does not change the mass of dust grains,
and accordingly does not change the number of dust grains in a given
mass interval. However, shattering leads to the transfer of dust grains
from a given bin to bins with different aromatization degrees.

\section{CONTRIBUTION OF VARIOUS FACTORS TO AROMATIZATION AND
DESTRUCTION OF DUST}

In this section, we present illustrative computations of variations of
the main characteristics of hydrocarbon dust grains under various conditions
characteristic for the ISM. Since our main interest is in the IR emission of
aromatic compounds, we begin with a computation of the aromatization of
dust grains under the action of UV radiation, which leads to the removal 
of hydrogen atoms. Figure~1 presents the evolution of the width of the
bandgap $E_{\textrm{gap}}$ for dust grains with various numbers of carbon
atoms $N_{\textrm{C}}$, located in radiation fields with various intensities.
We assume that the spectrum is a multiple the radiation field estimated
in [71] and use
a dimensionless coefficient $U$ equal to unity in the vicinity of the Sun
as a measure of radiation intensity. We did not take into account the
second stage of dissociation in these computations. Note that the shape of
the spectrum will be different in SFRs, which are a main potential target
for studies using this version of our model, and the model makes it possible
to take this into account. However, in this study, we used a
parametrization of the spectrum for the vicinity of the Sun for illustrative
purposes, and also to enable comparison with the results of previous studies.

\begin{figure}[t!]
\includegraphics[width=0.8\textwidth]{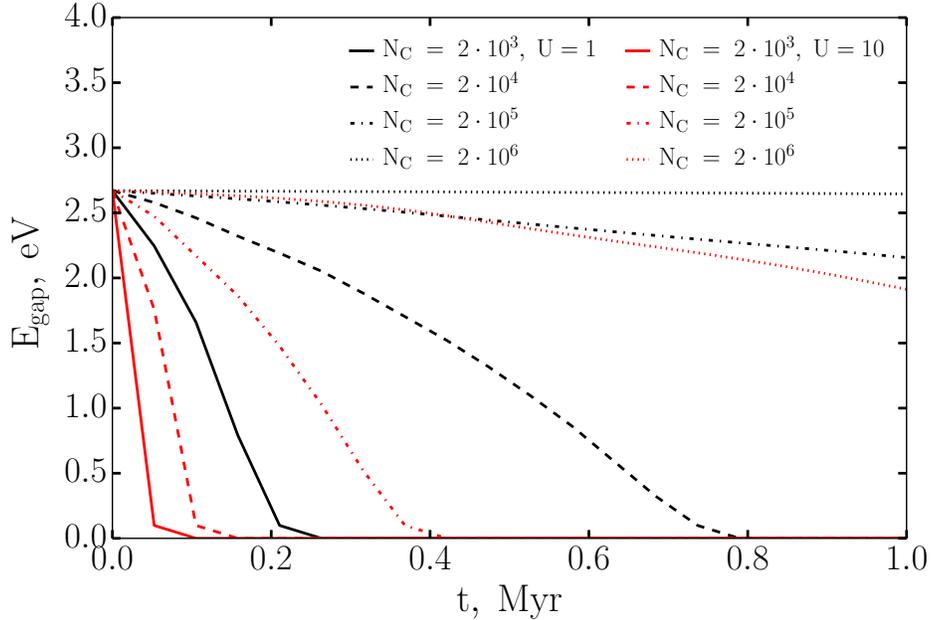}
\setcaptionmargin{5mm}
\onelinecaptionsfalse
\caption{Variation of the width of the bandgap $E_{\textrm{gap}}$ under
the action of a radiation field for \hbox{a-C:H} dust grains with various
numbers of carbon atoms $N_{\textrm{C}}$. Results are presented for
radiation fields with $U=1$ (black) and $U=10$ (red).}
\end{figure}

Small dust grains with $N_{\textrm{C}}<2\times 10^4$, corresponding to radii
less than 35~\AA, lose essentially all their hydrogen atoms in
several hundred thousand years, even when the radiation field illuminating
them is not very strong ($U=1$). Consequently, there should be no small,
hydrogenated, carbon dust grains in the ``ordinary'' ISM, since they are
very rapidly transformed into aromatic grains. The restructuring of larger
dust grains takes longer: when $N_{\textrm{C}}\sim 2\times10^4$ (a radius of
about 75~\AA), the aromatization time is nearly a million years. However,
they are also most likely in a dehydrogenated state in the ISM. Dust grains with
$N_{\textrm{C}}>2\times10^5$ ($a>200$~\AA) evolve over long times, and
remain predominantly hydrogenated. Increasing the strength of the radiation
field by a factor of ten ($U=10$) allows for appreciable
aromatization of larger dust grains on time scales less than a million years. 
Such radiation fields (and more intense fields) are characteristic, for
example, for HII complexes, whose lifetimes comprise several million years.
Thus, large dust grains in these objects could be partially aromatized.

In Figure~2 we compare the probabilities for removal of a \hbox{C$_2$}\ group
for various ionization probability functions and for dust grains with various
numbers of carbon atoms. The plots for $Y_{\textrm{ion}}^{\textrm{V}}$ 
are shown in red, and those for $Y_{\textrm{ion}}^{\textrm{WD}}$ are shown in black.
When the function $Y_{\textrm{ion}}^{\textrm{WD}}$ is used, dust grains with
16 and 48 atoms are efficiently destroyed at photon energies more than ${\sim}15$~eV, 
while the destruction of dust grains with 128 atoms begins only for photon
energies greater than 17~eV. When the function $Y_{\textrm{ion}}^{\textrm{V}}$
is used, which assumes that the energy of the absorbed hard photons goes
completely into ionization, the destruction of dust grains with 16 or more 
atoms becomes inefficient at photon energies of about 17~eV, which seems unrealistic.
Moreover, this assumption contradicts recently published experimental
results~[72], according to which ionization dominates for photon energies
from 8 to 40~eV only for large dust grains, while a more important channel
for small dust grains is their dissociation (loss of a hydrogen atom).

\begin{figure}[t!]
\includegraphics[width=0.8\textwidth]{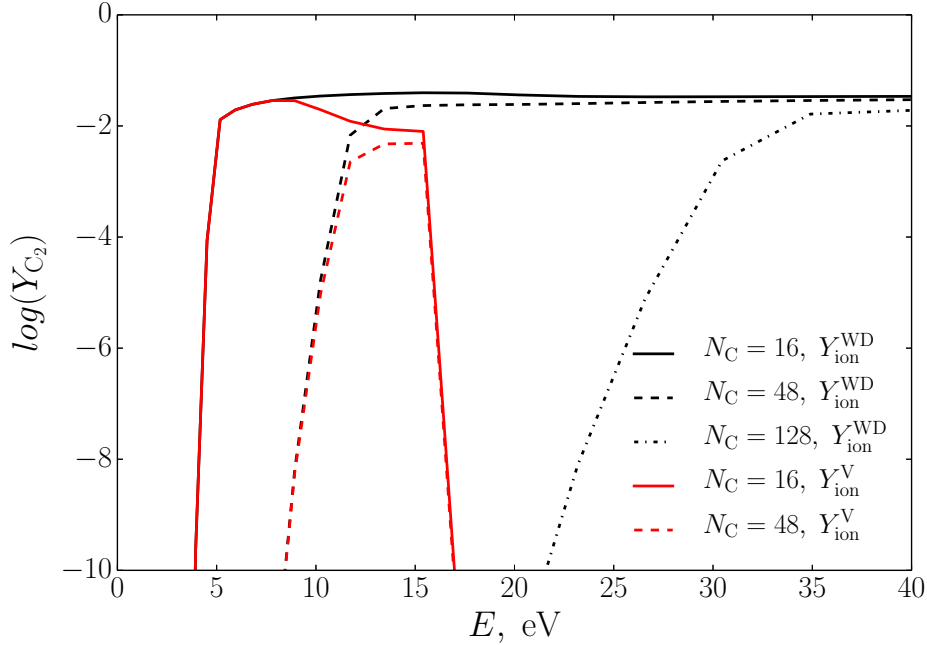}
\setcaptionmargin{5mm}
\onelinecaptionsfalse
\caption{Dependence of the probability of destruction of dust grains with
numbers of carbon atoms 16, 48, and 128 for the ionization probability
functions $Y_{\textrm{ion}}^{\textrm{WD}}$ (black) and
$Y_{\textrm{ion}}^{\textrm{V}}$ (red).}
\end{figure}

As was noted above, at the second stage of photodissociation the rate of
destruction of dust grains strongly depends on the parameter $E_0$. Figure~3
shows the dependence of the probability for the removal of a \hbox{C$_2$}\ 
group ($Y_{\textrm{C}_2}$) on $E_0$ for dust grains with $N_{\textrm{C}} = 
16$, 28 and 48 that absorb photons with energies of 8 and 12~eV. The
probability of dissociation falls off strongly with increasing $E_0$. 
For the same dust grain that absorbs a photon with energy 12~eV, the 
probabilities of losing a \hbox{C$_2$}\ group for $E_0=3$ and 5~eV differ
by five orders of magnitude. It is obvious that computations of the
evolution of ensembles of dust particles in SFRs must consider various 
values for this parameter.

\begin{figure}[t!]
\includegraphics[width=0.8\textwidth]{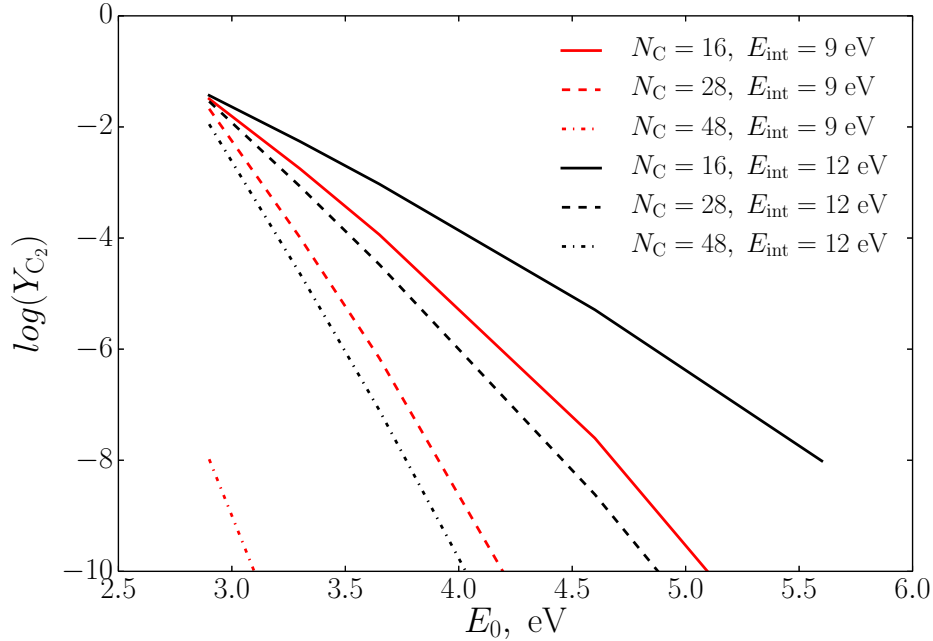}
\setcaptionmargin{5mm}
\onelinecaptionsfalse
\caption{Dependence of the probability of removing a \hbox{C$_2$}\ group
on the parameter $E_0$ characterizing the energy for breaking the C--C bonds.
Computations are shown for dust grains with numbers of carbon atoms 16, 28, 
and 48 (solid, dashed, and dot-dashed lines, respectively) for internal
energies of 9~eV (red) and 12~eV (black). \hfill}
\end{figure}

The rate of destruction of \hbox{a-C:H} dust grains calculated using 
formula~(5) is shown in Figure~4 for $E_0=2.9$~eV (black) and  $E_0= 5$~eV
(red). As $E_0$ is decreased, a grain with $N_{\textrm{C}}\sim 50$ 
can be destroyed by a radiation field with $U=1$ at a rate of about
$10^{-11}$ carbon atoms per second; i.e., about $3\times10^{-4}$ \hbox{C$_2$}
groups are dissociated in a year, and the entire dust grain will be
destroyed over about 80\,000~yrs. However, the destruction rate falls
rapidly with increase in the number of atoms. For example the complete
destruction of dust grains with $N_{\textrm{C}}\sim 60$ requires about
10~million years. Thus, when $E_0=2.9$~eV, dust grains with more than
50 atoms are stable, and can survive over the characteristic lifetime of
an HII complex. If we adopt higher values for $E_0$ (5~eV), photodestruction
becomes unimportant. Even molecules with 20 atoms will be destroyed in a
radiation field with $U=100$ only after a million years. Such high values
of $U$ are not generally characteristic of HII complexes, but the intensities
of the radiation fields inside individual HII regions can exceed the
radiation field in the solar vicinity by a factor of tens or hundreds
of thousands; photons beyond the Lyman limit are also present in HII regions,
suggesting that photodestruction can also play a key role in these objects
with higher values of $E_0$.

\begin{figure}[t!]
\includegraphics[width=0.8\textwidth]{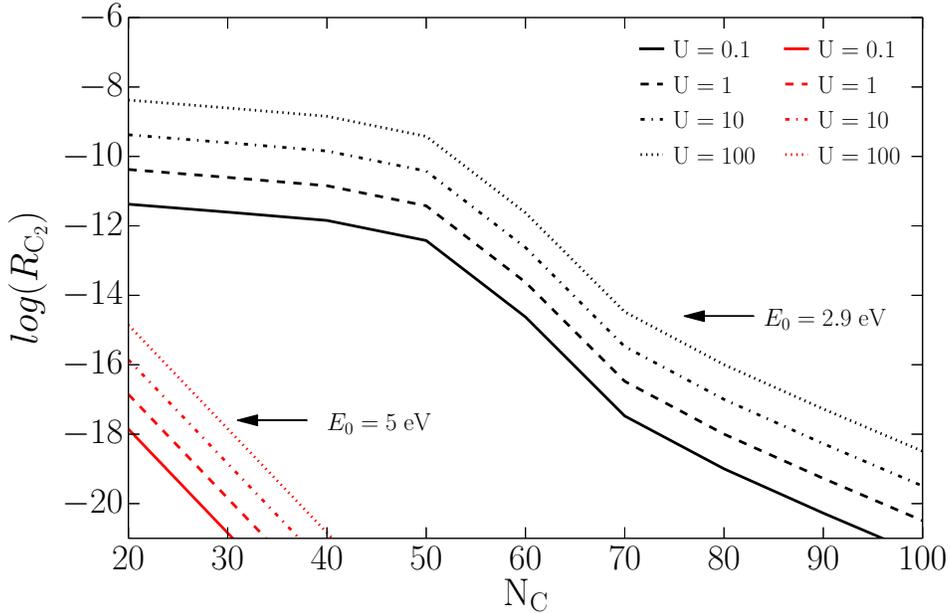}
\setcaptionmargin{5mm}
\onelinecaptionsfalse
\caption{Dependence of the rate of photodestruction of dust grains via the
removal of a \hbox{C$_2$}\ group on the number of carbon atoms. Results
are presented for radiation fields with $U=0.1$, $1$, $10$, and $100$ 
and for $E_0=2.9$~eV (black) and 5~eV (red).}
\end{figure}

These estimates are in agreement with the results of other studies. For
example, Allain et al.~[47] concluded that photodestruction plays a role
in the circumsolar radiation field only for PAHs with fewer than 50 atoms.
The critical number of carbon atoms estimated in~[56] is about 30--40. 
Our results are in agreement with these estimates, but we must recall that
the critical number of atoms depends strongly on the adopted value of
the dissociation energy $E_0$.

\begin{figure}[t!]
\includegraphics[width=0.4\textwidth]{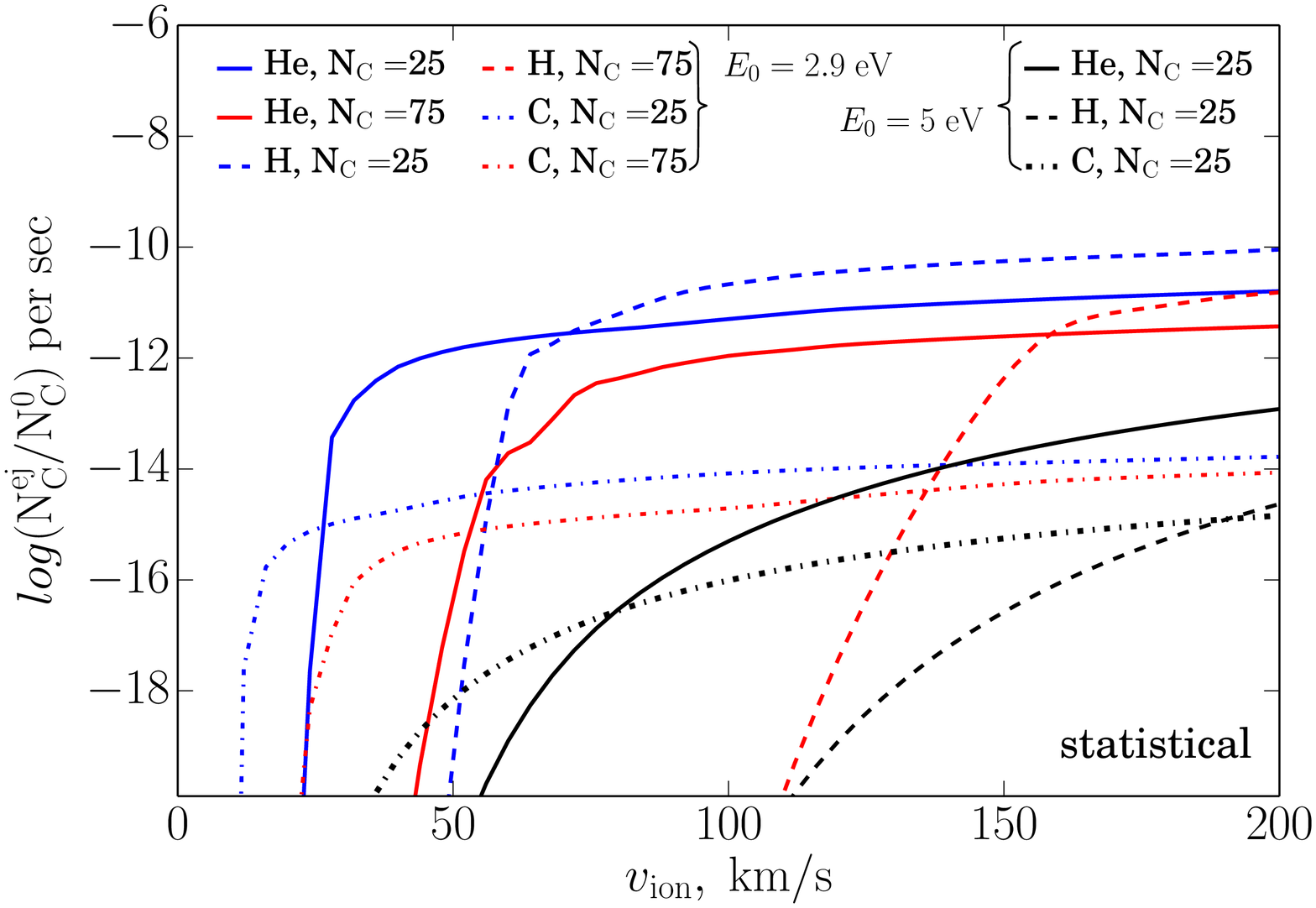}
\includegraphics[width=0.4\textwidth]{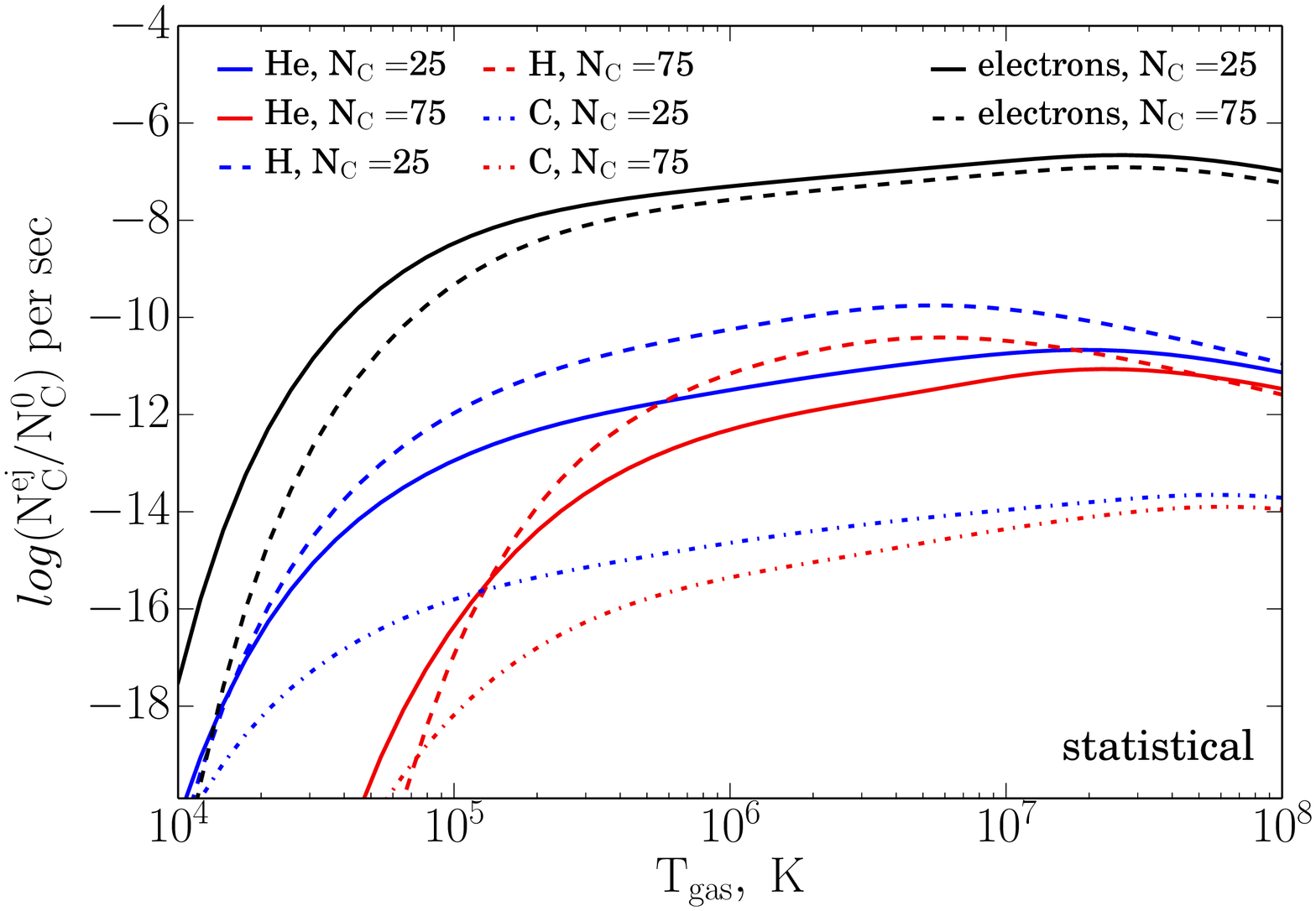}
\setcaptionmargin{5mm}
\onelinecaptionsfalse 
\caption{Dependence of the rates of statistical sputtering of dust grains on
the velocity of a collision with non-thermal ions (left) and the gas
temperature (right). Left: collision of a dust grain with $N_{\textrm{C}}=25$ 
and an ion of hydrogen (blue dotted), helium (blue solid), and carbon
(blue dot-dashed) with $E_0=2.9$~eV. The corresponding set of lines in
red shows the analogous results for a collision with a dust grain with
$N_{\textrm{C}}=75$, also for $E_0=2.9$~eV. The corresponding set of lines
in black shows the analogous results for a collision of a dust grain with
$N_{\textrm{C}}=25$, but for $E_0=5$~eV. Right: blue and red curves
show results for collisions of the same dust grains and ions for  
$E_0=2.9$~eV, but for thermal ions. The black solid and dashed curves
denote the rates of destruction of dust grains with $N_{\textrm{C}}=25$ and
75, respectively, by thermal electrons.}
\end{figure}

The role of statistical disruption by non-thermal (left) and thermal (right)
particles is illustrated in Figure~5, which presents the computational results
in the form of relative decrease in the number of carbon atoms in a dust
grain per second. Ions of the most abundance elements --- hydrogen and
helium --- are considered, as well as carbon ions as a representative of 
the population of heavier ions. These computations assumed the following
number densities:
$n_{\textrm{H}^{+}}=1$~cm$^{-3}$,
$n_{\textrm{He}^{+}}=0.1$~cm$^{-3}$, and
$n_{\textrm{C}^{+}}=10^{-4}$~cm$^{-3}$. It is obvious that with these
number densities and $E_0=2.9$~eV, collisions at velocities lower than
${\sim}20$~km/s are not sufficient to destroy any dust grains. Formally,
carbon ions dominate at these velocities, but the relative decrease in
atoms over $10^6$~yrs does not exceed $0.1$, even for the smallest
dust grains.

The destruction of small dust grains by helium ions dominates at velocities
from 20 to 50~km/s, leading to a relative decrease in the number of atoms
of order $10^{-1}$ over $10^3$~yrs. At velocities above 50~km/s, small
dust grains are destroyed mainly by hydrogen ions. Increasing the velocity
of collisions with H$^{+}$ ions to 100~km/s leads to the destruction of
dust grains with $N_{\textrm{C}}=25$ over less than $10^3$~yrs. Taking into
account the fact the enhanced densities in shock fronts, where 
$n_{\textrm{H}}$ can reach $10^4$~cm$^{-3}$, the destruction time can be
shortened to ${\sim}0.1$~year. The rate of destruction of dust grains with
large numbers of atoms ($N_{\textrm{C}}=75$) is approximately an order of
magnitude lower than the rate for dust grains with $N_{\textrm{C}}=25$.

If we adopt $E_0=5.0$~eV, the minimum velocity required for destruction,
$v_{\textrm{ion}}^0$, increases by about a factor of two, while the 
destruction rate decreases by several orders of magnitude. For example,
with a collision velocity of 100~km/s, the destruction of dust grains with
$N_{\textrm{C}}=25$ requires no less than $10^7$~yrs at the given
number density, with the main role in the sputtering of dust grains 
being played by helium ions.

Figure~5 (right) shows the destruction of dust grains by thermal ions and
electrons. Destruction by electrons dominates over destruction by ions at
all the temperatures considered. When the temperature of the ambient gas
is $10^5$~K or more, dust grains with both 25 and 75 carbon atoms are
destroyed in about a year. The rate of destruction by 
thermal ions of a gas at such temperatures is comparable with the rates of destruction
by non-thermal ions with collision velocities of order 100~km/s. Non-thermal
ions of hydrogen and helium play an important role in the destruction of
dust only if the gas temperature does not exceed $10^4$~K. At such high
temperatures, the dust grains are predominantly destroyed by collisions with
thermal electrons. This type of destruction is most important in supernova
remnants and, more general, in the hot phase of the ISM. The turbulent velocity
dispersion in SFRs does not exceed several tens of km/s,
and the maximum temperatures near massive stars can reach several thousand
Kelvin. Under these conditions, statistical sputtering will always be
less efficient than photodestruction. 

Figure~6 presents the results of our computations of the rate of destruction
of dust grains by non-statistical sputtering. At low velocities (to
70~km/s), the main role is played by helium ions, while collisions with
hydrogen ions are also important at higher velocities. Due to their lower
abundance, carbon ions play virtually no role. The rates of destruction of
dust grains with 25 and 75 atoms are virtually the same, in contrast to the
case of statistical destruction; at high velocities, the destruction rate
is even higher for larger dust grains. This is expected because the cross section
for interaction (the cross section of the dust grains) increases with their
size, raising the probability of an incident particle collision with
one of the atoms in the lattice.

\begin{figure}[t!]
\includegraphics[width=0.4\textwidth]{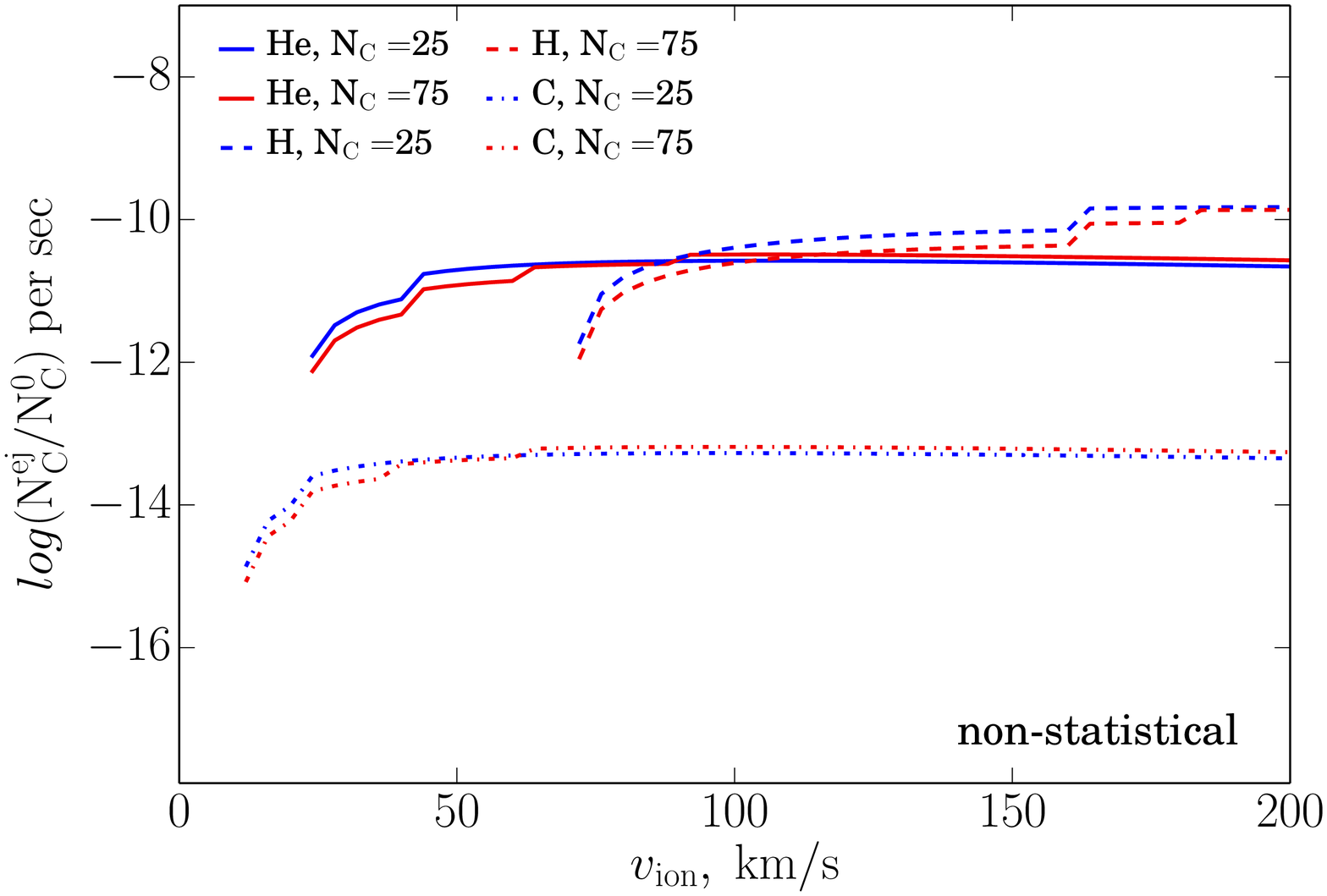}
\includegraphics[width=0.4\textwidth]{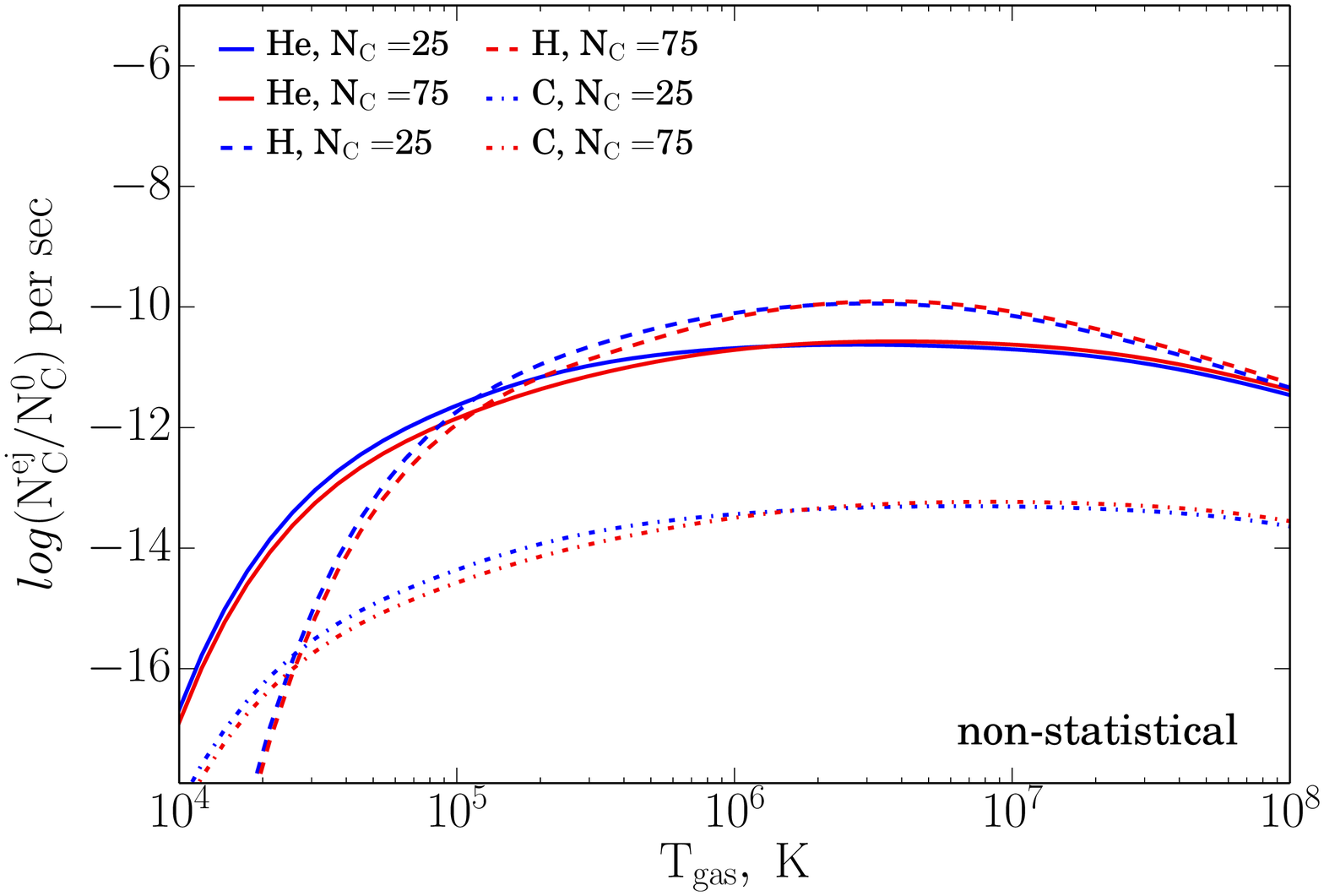}
\setcaptionmargin{5mm}
\onelinecaptionsfalse
\caption{Dependence of the rates of non-statistical sputtering of dust grains
on the velocity of collisions with non-thermal ions (left) and on the gas
temperature (right). Left: collisions of a dust grain with 
$N_{\textrm{C}}=25$ and ions of hydrogen (blue dashed), helium (blue solid),
and carbon (blue dot--dashed). The corresponding set of lines in red
shows the analogous results for collisions of dust grains with 
$N_{\textrm{C}}=75$. Right: the blue and red curves denote collisions of
the same dust grains and thermal ions.}
\end{figure}

In contrast to statistical destruction, the rates of this interaction are
significant at low velocities: helium ions are capable of destroying a grain
with $N_{\textrm{C}}\le75$ after $10^3$~yrs, even with collision velocities
of the order of 50~km/s. At higher velocities, the destruction rates by
ions of hydrogen and helium for elastic and inelastic interactions become
comparable. This is also true for thermal ions: at temperature above
$10^5$~K, the destruction rate by ions is several orders of magnitude lower
than the rate of statistical destruction by thermal electrons.

\begin{figure}[t!]
\includegraphics[width=0.8\textwidth]{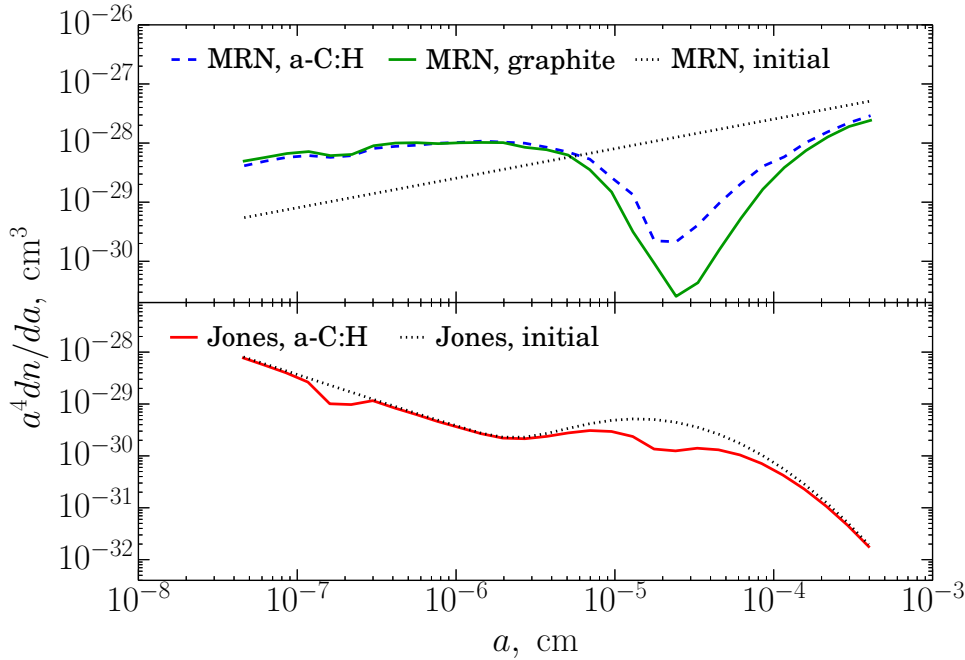}
\setcaptionmargin{5mm}
\onelinecaptionsfalse 
\caption{Upper: size distribution of dust grains up to the onset of
shattering (black dotted curve) and after one million years, for an MRN
distribution. The green curve shows the distribution for graphite grains,
and the blue curve the distribution for \hbox{a-C:H} dust grains. Lower:
same plot for the initial size distribution taken from~[4] and \hbox{a-C:H}
dust grains. \hfill}
\end{figure}

Finally, let us consider the modeling results for the shattering of
dust grains. Figure~7 (top) shows the evolution of the size distribution 
of the dust grains for an initial MRN distribution~[73], extended toward
small dust grains to a radius of 4~\AA\ and toward large dust grains to
a radius of 0.4~$\mu$m, as was done in~[41]. The velocity dispersion of
the dust grains was taken from~[74] for the case corresponding to a thermal
ionized medium. For comparison, in addition to amorphous \hbox{a-C:H} dust 
grains, the upper panel in Figure~7 also shows the evolution of the size
distribution for graphite dust grains. The shattering parameters for
these materials are different, and the redistribution of the dust-grain sizes
for the two types of grains proceeds differently. The reduction in the
number of large dust grains is more efficient for graphite, but this does
not significantly affect the increase in the number of small dust grains.

Figure~7 (bottom) presents an analogous computation for an initial size
distribution of the dust grains taken from~[4]. The redistribution of
sizes is less efficient in this case. In particular, instead of a growth
in the number of small dust grains, to some extent, the destruction of
dust grains of all sizes is observed. This occurs mainly due to the
initially high abundance of small dust grains, whose collisions lead to
their total destruction. There are more large than small dust grains in
the MRN distribution, so that the destruction of small dust grains is
compensated by the shattering of large dust grains. The size distribution
from~[4] is characterized by the opposite relationship between the numbers
of large and small grains, so that the destruction of small dust grains
is not balanced by the appearance of fragments of large dust grains.
Thus, the chosen size distribution is an important factor in modeling the
evolution of dust, and determines the trend in the variation of the
abundance of small dust grains.

\section{DISCUSSION AND CONCLUSION}

Problems related to the destruction of dust grains of various sizes
and chemical compositions have been discussed many times in the literature.
The novelty of the model presented here is that we have considered not only
the destruction of hydrocarbon dust grains of various sizes, but also 
variations in their structural properties. This latter element is important
for modeling the IR radiation of dust grains in the ISM and SFRS.  Let us 
briefly summarize the main characteristics of our model.  In the version 
presented here, we have assumed that hydrocarbon dust grains in SFRs initially
(after their formation in asymptotic giant branch stars or molecular clouds)
have predominantly aliphatic structure. Interactions of dust grains with
UV radiation and energetic particles in the ambient gas leads to a gradual
loss of hydrogen atoms and, thus, to the formation of aromatic bonds, either
in the entire volume of the dust grain (if its radius is less than 200~\AA), 
or in a surface layer with a thickness of 200~\AA.

We have considered each of these processes in detail, and have computed
the rates of the destruction (or restructuring) of dust for various parameters.
We have shown that the aromatization time for dust grains with
$N_{\textrm{C}}<2\times10^3$ is less than $2\times10^5$~yrs, even for a
moderate interstellar radiation field ($U=1$), so that small dust grains
should be fully aromatized in the ISM.

When the fraction of hydrogen atoms in a grain becomes less than $5\%$, 
the loss of \hbox{C$_2$}, and possibly C$_2$H$_2$, molecules is added to the
removal of hydrogen atoms. As a result, such grains are not only aromatized, 
but also destroyed. For characteristic evolutionary times in SFRs, the
second phase occurs only for small dust grains with less than 50 carbon
atoms. Larger dust grains will be affected by this process only in the
vicinities of hot massive stars.

Other destruction mechanisms associated with collisions of dust grains and
energetic gas particles are efficient only in the case of high collision
velocities and/or high gas temperatures. At temperatures above $10^5$~K,
the main mechanism for the destruction of small dust grains becomes
inelastic interactions with thermal electrons.

Finally, an important process leading to a redistribution of the dust-grain
sizes is shattering as a result of collisions between dust grains. 
This process is the main channel for compensating the destruction of
small dust due to interactions with photons and sputtering.

We have shown that the evolution of hydrocarbon dust grains of various
sizes in the ISM is expressed through variations of their size
distribution and structural properties, which should lead to variations
in their optical properties~[45, 46]. These variations will inevitably
be reflected in the emission and absorption properties of the dust component
of the ISM. Important applications of the model we have presented include
studies of variations of the ratio of the intensities of the emission bands
at 3.3 $\mu$m and 3.4 $\mu$, which may reflect the relative abundance of
aromatic and aliphatic hydrocarbons, and explaining the metallicity 
dependence of the ratio of the 8 and 24 $\mu$m fluxes. Another example of
an observational manifestation of the evolution of hydrocarbon dust that
we have considered here could be variations in the UV peak in the 
interstellar extinction curve~[75], which is often associated with 
differences in the number of small hydrocarbon dust grains. Our model can
be used to describe the evolution of the dust component not only of the
ISM and large star-forming complexes, but also in so-called infrared
bubbles observed around HII regions, and also in planetary nebulae and
supernova remnants. The relationship between the evolution of dust in the
ISM and its size distribution and the properties of its IR emission will
be considered in a future paper.

Our described model for the evolution of dust is not the only one possible.
For example, Chiar et al.~[34] have considered another sequence in which,
after their formation in evolved stars, hydrocarbon dust grains are 
predominantly aromatic, and form aliphatic mantles in the ISM as a result
of bombardment by H atoms. It was noted in~[76] that this is possible in
molecular clouds, where aliphatic mantles grow on dust grains due to the
accretion of hydrogen atoms. However, in diffuse media, we should observe
the situation we have considered --- large dust grains possessing an 
aliphatic core surrounded by an aromatic envelope and fully aromatic small 
dust grains.

\section{ACKNOWLEDGEMENTS}

We thank the referee for valuable comments and corrections, E.
Micelotta for consultations about the sputtering of PAHs, and V.
Akimkin for useful advice and his interest in these studies.
This work was partially supported by the Russian Foundation for Basic
Research (grants 13-02-00640, 14-02-31456, 15-02-06204), a grant
of the President of the Russian Federation (MK-4536.2015.2), the
``Dinastiya'' Foundation, and the Italian Ministero dell'Universita 
e della Ricerca.

\begin{flushright}
\emph{APPENDIX A}
\end{flushright}

\section*{COMPUTATION OF THE ENERGY TRANSFERRED TO A DUST GRAIN BY AN ION
DUE TO FRICTION}

The model assumes that an ion transfers an amount of energy to a 
\hbox{a-C:H} dust grain as it traverses one molecular layer. This energy
can be estimated using the formula propose in [35] for molecular PAHs:
\begin{gather*}\tag{\mbox{A}1}
E_{\textrm{tr}}^0 = \int\limits_{-s/2}^{s/2}
\gamma(r_{s})v_{\textrm{ion}} ds,
\end{gather*}
where $s$ is the distance along the trajectory of a photon as a function
of the angle $\theta$ between the layer and the ion trajectory:
\begin{gather*}\tag{\mbox{A}2}
s =
\begin{cases}
d/2\cos\theta \quad \theta < \textrm{arcctg}(d/2a) \\
a/2\sin \theta, \quad \theta > \textrm{arcctg}(d/2a) \\
\sqrt{d^2+a^2}, \quad \theta = \textrm{arcctg}(d/2a)
\end{cases}
\label{distance}
\end{gather*}
Here, $d$ and $a$ are the thickness (4.31~\AA~[97]) and radius of a single
layer of the molecular structure of the dust grains. The zero point is
located at the center of the PAH, or, in our case, in the center of the
layer.

The parameter $r_{s}$ depends on the electron density $n_0$ as 
$(4/3\pi n_0)^{-1/3}$, and the electron density depends on the configuration
of the molecules. It was proposed in [35] to use the following expression
for the electron density, based on the computations of~[95] for fullerene,
but taking into account the different shape of PAHs:
\begin{gather*}\tag{\mbox{A}3}
n_0 = 0.15 \exp[-(s \cos \theta)^2/2.7]. \label{pah_elect_dens}
\end{gather*}
The parameter $\gamma$ is calculated in terms of $r_{s}$:
\begin{gather*}\tag{\mbox{A}4}
\gamma(r_s) = \Gamma_{0}^{\textrm{ion}}
\exp(-(r_s-1.5)/R_2^{\textrm{ion}}) \label{fric_coef}
\end{gather*}
with $R_2 = (2.28, 0.88, 0.90)$ and $\Gamma_{0}^{\textrm{ion}}= (0.310, 
1.112, 0.690)$ for ions of hydrogen, helium, and carbon, respectively.

\begin{flushright}
\emph{APPENDIX B}
\end{flushright}

\section*{COMPUTATION OF THE ENERGY TRANSFERRED TO AN ATOM IN THE LATTICE
DUE TO A COLLISION WITH AN INCIDENT PARTICLE}

The maximum amount of energy $E_{\textrm{tr}}^{\textrm{max}}$ that an ion
can transfer in a head-on collision with a carbon atom in the lattice is
related to the kinetic energy of the ion through the coefficient $\gamma$, 
which can be calculated in terms of the masses of the colliding particles:
\begin{gather*}\tag{\mbox{B}1}
E_{\textrm{tr}}^{\textrm{max}} = \gamma E =
\frac{4M_{\textrm{ion}} M_{\textrm{C}}}{(M_{\textrm{ion}}
+M_{\textrm{C}})^2}E_0^{\textrm{ion}},
\end{gather*}
where $M_{\textrm{ion}}$ is the mass of the ion and $M_{\textrm{C}}$ the
mass of the carbon atom. Based on this expression, the minimum kinetic
energy of an incident particle in order for it to be able to knock out an
atom from the lattice with $T_{\textrm{crit}} = 7.5$~eV should be 26.8,
6 and 7.5~eV for H$^{+}$, He$^{+}$, and C$^{+}$ ions, respectively.
According to~[25], the interaction cross section $\sigma(E)$ per atom for
ions with energies above the critical energy can be calculated as follows:
\begin{gather}\tag{\mbox{B}2}
\sigma(E_0^{\textrm{ion}}) =
4 \pi a^{\textrm{scr}} Z_{\textrm{ion}} Z_{\textrm{C}}
e^2 \frac{M_{\textrm{ion}}}{M_{\textrm{ion}}+
M_{\textrm{C}}}  s_n \times\\
\nonumber{}\times \frac{1-m}{m}  \frac{1}{\gamma
E_0^{\textrm{ion}}}  \left(\left[
\frac{E_{\textrm{crit}}^{\textrm{ion}}}
{E_0^{\textrm{ion}}}\right]^{-m}-1\right)
\end{gather}
where  $a^{\textrm{scr}}$ is the screening length, $Z_{\textrm{ion}}$ 
and $Z_{\textrm{C}}$ are the charge numbers of the ion and carbon, 
$e$ is the electron charge, $s_n$ is the specific cross section for nuclear
deceleration, and $m$ is a parameter that depends on the energy through
a relation taken from~[69]. The mean energy transferred to the carbon
atom is given by
\begin{gather*}\tag{\mbox{B}3}
\langle E_{\textrm{tr}}^{\textrm{elas}}\rangle =
\frac{m}{1-m}\gamma
\frac{(E_0^{\textrm{ion}})^{1-m}-(E_{\textrm{crit}}^{\textrm{ion}})^{1-m}}{(E_{\textrm{crit}}^{\textrm{ion}})^{-m}-(E_{\textrm{tr}}^{\textrm{max}})^{-m}}
\end{gather*}

\end{document}